\documentclass[seceq,preprint]{ptptex}
\usepackage{graphicx}
 \notypesetlogo
  \def\bea*{\begin{eqnarray*}}
 \def\eea*{\end{eqnarray*}}
\def\ind{\indent}
 \def\nn{\nonumber}
 \def\be{\begin{equation}}
 \def\ee{\end{equation}}
 \def\hx{{\hat x}}
 \def\hS{{\hat S}}
 \def\btheta{{\bar \theta}}
 \def\htheta{{\hat \theta}}
 
 \def\beq{\begin{eqnarray}}
 \def\eeq{\end{eqnarray}}
 \def\ba{\begin{array}}
 \def\ea{\end{array}}
 \def\cL{{\mbox {${\cal L}$}}}

 \def\cL{{\mbox {${\cal L}$}}}
 
 \def\vs{\vspace{5mm}}
  \def\lslash{l{\raise 0.8pt\hbox{$\!\!\!/$}}}
  \def\qslash{q{\raise 0.8pt\hbox{$\!\!\!/$}}}
  \def\pslash{p{\raise 0.8pt\hbox{$\!\!\!/$}}}
  \def\rslash{r{\raise 0.8pt\hbox{$\!\!\!/$}}}
  \def\trslash{{\tilde r}{\raise 0.8pt\hbox{$\!\!\!/$}}}
  \def\sslash{s{\raise 0.8pt\hbox{$\!\!\!/$}}}
\title{%
Noncommutative Regularization In Gauge Theories
}
\vspace{5mm}
\author{%
  Katsusada {\sc Morita}
}
\inst{%
{\it  Department of Physics, 
  Nagoya University, Nagoya 464-8602, Japan}
}
\vspace{5mm}
\abst{%
Gauge invariance of noncommutative (NC) regularization which, on the basis
of a Lorentz-invariant NC action regarded as a `regulated' action, neither 
introduces auxiliary fields nor extends dimensions to complex values, is 
proved by explicitly calculating photon self-energy in the one-loop
approximation in scalar QED. Transversality of vacuum polarization in NC 
regularization is also briefly reviewed comparing with Pauli-Villars-Gupta
and dimensional regularizations. NC regularization is applied to 
gauge-invariant calculation of one-loop gluon self-energy in $U(N)$ gauge 
theory. It is shown that $U(1)$ decouples from $SU(N)$ in the one-loop 
gluon self-energy diagrams. That is, gauge-invariant result on the one-loop 
$SU(N)$ gluon self-energy is obtained from consideration of 
Lorentz-invariant NC $U(N)$ gauge theory.
}
\begin{document}
\maketitle
\baselineskip=17.8pt
%
%%%%%%%%%%%%%%%%%%%%%%%%%%%%%%%%%%%%%%%%%%%%%%%%%%%%%%%%%%%%%%%%%
\section{Introduction}
%%%%%%%%%%%%%%%%%%%%%%%%%%%%%%%%%%%%%%%%%%%%%%%%%%%%%%%%%%%%%%%%%
To handle UV divergences in quantum field theory (QFT) to carry through 
renormalization it is necessary to `regulate' Feynman amplitudes
in a way compatible with Ward-Takahashi identities. In spite of its purely 
technical nature any gauge-invariant regularization of divergent integrals 
is necessary mathematical device of obtaining sensible physical result
in perturbative QFT. 
\\
\ind
Among many regularization techniques Pauli-Villars-Gupta and dimensional 
regularizations are well-known. In the former one first introduces some 
(normal and abnormal) auxiliary fields in the Lagrangian density, 
obtaining `regulated' action. Their quanta are given infinitely large 
masses to be unobservable. The minimum number of the auxiliary fields 
depend on the model. On the other hand, dimensional regularization 
defines Feynman amplitudes as analytic functions in complex space-time 
dimension $n$ so that only physical particles run through internal loops, 
assuming them to propagate in complex dimensional space-time. Divergences 
appear as poles at $n=4$ and/or $n=2$. Physical results are obtained after 
subtraction thereof thanks to gauge invariance. Since dimensional 
regularization is especially convenient for non-Abelian gauge theory on 
the basis of which the standard model is constructed, it becomes 
indispensable for perturbational calculations in QFT and is now widely 
used in the literature.
\\
\ind
In comparison with them noncommutative (NC) regularization we have recently
proposed\cite{1)} deals with only physical fields and keeps dimensions 4, 
yet possible to `regulate' Feynman amplitudes in a gauge-invariant way.
The mechanism of regularization is quite different. One first computes
{\it finite} amplitudes based on Lorentz-invariant NC action. They contain, 
however, IR singularity in Euclidean metric, which is a necessary consequence
of recovering QFT in the commutative limit. The presence of IR singularity
brings about a new problem upon continuation back to Minkowski metric,
which is avoided only if consistent `subtraction' is carried out. The 
`subtraction' reproduces the well-known renormalized amplitudes. We would 
like to explain what motivated us to formulate NC regularization.
Before doing it we have to confess that whether or not it works in multi-loops
and even one-loop with three and four vertices has yet to be investigated.
\\
\ind
Quantum field theory on NC space-time (NCQFT)\cite{2)} has been investigated 
extensively in recent years. The upsurge is revived by Seiberg 
and Witten\cite{3)} who realized that, when open strings propagate 
under constant background $B$ field, the coordinates they attach on 
$D$-branes become noncommutative. There is another strong motivation that 
space-time noncommutativity at, say, Planck scale provides a fascinating 
possibility of modifying the conception of the structure of space-time, 
which may shed light on the long-standing divergence problem in QFT. One then 
naturally expects that NCQFT, if consistently formulated, would suggest a step 
forward beyond standard picture of present-day particle theory. The purpose of
the present paper is to convey,  against the current dominant streams in the 
study of NCQFT, our biased view that the lack of Lorentz symmetry in NCQFT may 
be a fundamental obstacle to go beyond (relativistic) QFT. If Lorentz symmetry 
is restored {\it without} encountering singularity, a finite theory would be 
dreamed.
\\
\ind
It is well-known that NCQFT violates Lorentz symmetry. This is apparent because 
NC parameter $\theta^{\mu\nu}$ defined by
\be
[\hx^\mu,\hx^\nu]=i\theta^{\mu\nu},\quad
\quad\mu,\nu=0,1,2,3,
\label{eqn:1-1}
\ee
is a constant anti-symmetric matrix, singling out one particular inertial frame 
from others. Here, $\hx^\mu (\mu=0,1,2,3)$ are the space-time coordinates 
represented by hermitian operators, which are assumed to transform as 
4-vector under the Lorentz transformations. This assumption implies that 
one can define Lorentz-covariant fields to describe interactions on NC space
(\ref{eqn:1-1}). There is no problem in the tree level if one accepts 
unavoidable appearance of Lorentz-violating parameters in observable quantities. 
Consideration of quantum effects changes the situation drastically. 
From numerous works\cite{2)} on NCQFT one learns that the Lorentz-violating 
parameter $\theta^{\mu\nu}$ causes unexpected features like IR/UV 
mixing\cite{4)} for nonplanar diagram and (consequent) unitarity 
problem.\cite{5)} In a sense they are pathological, but it is rather natural 
to suppose that the existence of IR/UV mixing implies that a commutative limit 
of NCQFT reproduces QFT with UV divergence provided that IR divergence can be 
isolated subject to {\it invariant} subtraction. This, in particular, means 
that IR singularity in perturbative NCQFT should be observed not only in 
nonplanar diagrams but also in planar diagrams in a Lorentz-invariant way so 
that the subtraction of IR singularity works as an equivalent alternative to 
the subtraction of UV divergence in QFT. 
\\
\ind
As a matter of fact, if Lorentz symmetry is assumed to stand as fundamental
in NCQFT as in QFT, it is no longer possible to consider $\theta^{\mu\nu}$
as constant. It should be regarded as an operator $\htheta^{\mu\nu}$. 
Lorentz-invariant NC space-time\footnote{Snyder\cite{6)} was the first
to introduce Lorentz-invariant NC space-time by assuming 
$\htheta^{\mu\nu}$ to be proportional to angular momentum operator.
We shall not consider this case because the associated momentum space is 
curved but not flat.}, called quantum space-time, intimately connected with 
NC space-time
(\ref{eqn:1-1}) was proposed ten years ago by Doplicher, Fredenhagen and 
Roberts (DFR)\cite{7)}. DFR assumed $\htheta^{\mu\nu}$ to be {\it central} 
so that the irreducible representations of the DFR algebra 
are characterized by an anti-symmetric second-rank tensor, 
$\theta^{\mu\nu}$, the eigenvalue of the operator $\htheta^{\mu\nu}$, and 
the algebra (\ref{eqn:1-1}) with tensorial $\theta^{\mu\nu}$ may be valid 
in a particular representation space of the DFR algebra. Feynman 
rules of QFT defined on quantum space-time are derived by Filk\cite{8)} 
who, within a single irreducible representation of the DFR algebra, found 
that UV divergence persists for planar diagram, while nonplanar diagram is 
regulated by the noncommutativity assumption. Minwalla, Raamsdonk and 
Seiberg\cite{4)} studied perturbation theory of NC scalar models and showed 
that such a regularization of nonplanar diagram generates IR singularity 
which would instead show up as UV divergence in the commutative limit. They 
termed the phenomenon IR/UV mixing. IR/UV mixing was found for nonplanar 
diagrams only. The presence of IR/UV mixing in perturbative NCQFT indicates 
that NCQFT correlates short-distance (UV) with long-distance (IR) behaviors 
in an intriguing way and makes it impossible for NCQFT to satisfy the 
correspondence principle in the sense that it possesses `classical' limit,
i.e., the commutative limit of NCQFT exists and should be identical to QFT.
This is simply because IR limit defined in Ref.~4) corresponds to the 
commutative limit so that IR singularity automatically excludes the 
existence of the commutative limit of NCQFT. This conclusion, which is also 
obtained by Hayakawa\cite{9)} for NC $U(1)$ gauge theory coupled to 
fermions (NCQED), is valid only for nonplanar diagrams because their 
formulation of IR/UV mixing did not meet Lorentz invariance: the result 
explicitly contains the 
Lorentz-violating parameters which affect loop integrals in nonplanar
but not planar diagrams. Only if one manages to `subtract off' IR 
singularity in an {\it invariant} way (as explained in the paragraph
containing (\ref{eqn:1-1})), can NCQFT 
go over to QFT in a {\it smooth} way in the commutative limit. In other 
words, we should yet look for NCQFT which satisfies the correspondence 
principle. Restricting to a particular representation space of the DFR 
algebra does not guarantee the validity of the correspondence principle.
\\
\ind
NCQFT without Lorentz violation proposed by Carlson, 
Carone and Zobin (CCZ)\cite{10)} is also based on the DFR algebra.
These authors treated NC parameter $\theta^{\mu\nu}$ as a kind of 
`internal' coordinates. This results in $\theta$-integration\footnote{This
amounts to take into account all irreducible representations of the DFR 
algebra.} of NC action that now contains fields defined on 10-dimensional 
space, 4 for the usual space-time and 6 for `internal' coordinates, 
$\theta^{\mu\nu}$. They asserted that only non-gauge theory allows fields 
not to `depend' on `internal' coordinates. In such a case we can apply 
perturbation theory. On the other hand, perturbation theory cannot be 
applied to gauge theory because it is impossible to determine vertices 
involving fields defined on the 10-dimensional space in terms of simple 
rules. In fact, CCZ resorted to the so-called $\theta$-expansion\cite{11)}
to calculate $S$-matrix element in their Lorentz-invariant NCQED.
\\
\ind
We took in I a different view point that perturbation theory can be 
applied to both non-gauge and gauge theories in Lorentz-invariant NCQFT.
We should then find IR singularity also in gauge theory since we already 
found\cite{12)} IR singularity in NC $\phi^4$ model in CCZ 
formalism.\footnote{Perturbative calculation was made possible because 
NC $\phi^4$ model is a non-gauge theory.} IR singularity depends on
external momenta but should be `subtracted off' so as to satisfy the 
correspondence principle. If otherwise, such Lorentz-invariant NCQFT 
does not make sense and should be put in the garbage 
bag. If, on the other hand, one succeeds in finding an invariant subtraction
method, the Lorentz-invariant NCQFT merely works to provide a `regulated' 
action. We could not then hear new physics from it. See, however, comments
on possible dual roles of the Lorentz-invariant NCQFT in the last section.
\\
\ind
Along this line of thought we carefully investigated\cite{1)} the unitarity 
problem\cite{5)} in NC $\phi^3$ model and vacuum polarization in 
Lorentz-invariant NCQED. Our Lorentz-invariant NC action is obtained by
integrating the conventional NC action over $\theta^{\mu\nu}$, 
\be
\hS
=\int\!d^{\,4}xd^{\,6}\theta\,
W(\theta)\cL(\varphi(x), \partial_\mu\varphi(x))_*.
\label{eqn:1-2}
\ee
Here, \footnote{$d^4x=dx^0dx^1dx^2dx^3, 
\;\;\;d^6\theta=
d\theta^{01}d\theta^{02}d\theta^{03}
d\theta^{23}d\theta^{31}d\theta^{12}$.
The conventional NC action $\int\!d^{\,4}x
\cL(\varphi(x), \partial_\mu\varphi(x))_*$
is popular.\cite{2)}} $\varphi(x)$ is a field variable
to be quantized, the subscript $\ast$ of the Lagrangian indicates that
the Moyal $\ast$-product  
\begin{eqnarray}
\varphi_1(x)*\varphi_2(x)\equiv
\varphi_1(x)e^{\frac i2
\theta^{\mu\nu}{\mbox{\scriptsize$
{\overleftarrow{{\partial_\mu}}}$}}
{\mbox{\scriptsize$\overrightarrow{\partial_\nu}$}}}
\varphi_2(x),
\label{eqn:1-3}
\end{eqnarray}
should be taken for all products of the field variables and $W(\theta)$ 
is a Lorentz-invariant weight function with the normalization\footnote{The
weight function $W(\theta)$ was first introduced in Ref. 10). It was 
later\cite{13)} found that there is a nuisance in the normalization 
condition and the moment formula in Ref. 10).}
\begin{eqnarray}
\int\!d^6\theta
W(\theta)=1.
\label{eqn:1-4}
\eeq
We define the length parameter $a$ by 
\begin{eqnarray}
\theta^{\mu\nu}=a^2{\bar\theta}^{\mu\nu}
\label{eqn:1-5}
\end{eqnarray}
with ${\bar\theta}^{\mu\nu}$ dimensionless. The commutative limit is 
obtained by taking the limit $a\to 0$. The normalization condition 
(\ref{eqn:1-4}) is independent of $a$,
\beq
W(\theta)&=&a^{-12}w({\bar\theta}),\nn\\[2mm]
\int\!d^6\theta
W(\theta)&=&\int\!d^6\btheta
w(\btheta)=1.
\label{eqn:1-6}
\eeq
It was shown\cite{1)} that the unitarity problem in NC $\phi^3$ model is 
caused by Lorentz violation and our Lorentz-invariant NC action avoids it, 
working as a `regulated' action. NC regularization takes a new UV limit\cite{12)} 
of Feynman amplitudes calculated based on (\ref{eqn:1-2}) such that
\begin{eqnarray}
\Lambda^2&\to&\infty,\qquad
a^2\to 0,\qquad
\Lambda^2 a^2:{\rm fixed}.
\label{eqn:1-7}
\end{eqnarray}
Here, $\Lambda$ denotes UV cutoff introduced to evade IR singularity. It is 
essential to realize that IR limit cannot be distinguishable from the
commutative limit which is characterized by a {\it single} Lorentz scalar. 
It is this feature coming from Lorentz invariance that the new UV limit 
works to eliminate IR singularity and, as a consequence, UV divergence 
from the theory.
\\
\ind
By calculating (one-loop) vacuum polarization in QED it was also 
shown\cite{1)} that 
the method preserves gauge invariance {\it without} cancellation. In the 
present paper we apply NC regularization method to scalar QED and $U(N)$ 
Yang-Mills gauge theory considering one-loop self-energy corrections of 
gauge boson (one-loop photon and gluon self-energies, respectively,) 
and prove the gauge invariance.
\\
\ind
The present paper is organized as follows. The next section is intended 
to illustrate NC regularization method by considering photon self-energy
in the one-loop approximation in scalar QED. The Maxwell sector of 
NCQED,\cite{14),9)} which looks like a non-Abelian gauge theory by the 
noncommutativity assumption, was investigated in I with the 
result that three-point vertices including ghost-ghost-photon coupling 
disappear by $\theta$-integration in (\ref{eqn:1-2}). Tadpole diagram 
arising from four-point vertex was also studied there and will be reconsidered 
in the end of the next section. Vacuum polarization in spinor QED is 
revisited in \S 3 to compare with Pauli-Villars-Gupta and dimensional 
regularizations. We present in \S 4 one-loop calculation of gluon 
self-energy in Lorentz-invariant NC $U(N)$ Yang-Mills and show that 
$U(1)$ decouples from $SU(N)$ in the new UV limit. We recall that 
Armoni\cite{15)} found that $U(1)$ does not decouple from $SU(N)$ in 
the conventional NC $U(N)$ Yang-Mills in the commutative limit. 
Our conclusion is in sharp contrast 
to that obtained in Ref.~14) due to $\theta$-integration, our imposition of 
Lorentz invariance. \S 5 is devoted to discussions. 
Some technical details are postponed to the Appendices.
%%%%%%%%%%%%%%%%%%%%%%%%%%%%%%%%%%%%%%%%%%%%%%%%%%%%%%%%%%
\section{One-loop photon self-energy in scalar QED} 
%%%%%%%%%%%%%%%%%%%%%%%%%%%%%%%%%%%%%%%%%%%%%%%%%%%%%%%%%%
In this section we illustrate NC regularization method in scalar QED. 
To this purpose we start with Lorentz-invariant NC 
action of scalar QED given by
\beq
\hS&=&\int\!d^4x\int\!d^6\theta W(\theta)
[(D_\mu\phi(x))^{\dag}*(D^\mu\phi(x))
-m^2\phi^{\dag}(x)*\phi(x)]
+\hS_{EM},
\label{eqn:2-1}
\eeq
where $\phi(x)$ is a complex scalar field subject to the $*$-gauge 
transformation, 
\beq
\phi(x)\to ^g\!\!\phi(x)&=&U(x)*\phi(x),\;\;\;
U(x)*U^{\dag}(x)=U^{\dag}(x)*U(x)=1,
\label{eqn:2-2}
\eeq
so that covariant derivative is defined by
\beq
D_\mu\phi(x)&=&\partial_\mu\phi(x)-ieA_\mu(x)*\phi(x),\nn\\[2mm]
A_\mu(x)&\to&^g\!A_\mu(x)=
U(x)*A_\mu(x)*U^{\dag}(x)+\frac ieU(x)*\partial_\mu U^{\dag}(x).
\label{eqn:2-3}
\eeq
The field strength tensor associated with $U(1)$ gauge field $A_\mu$ 
is defined by
\begin{eqnarray} 
F_{\mu\nu}(x)&=&\partial_\mu A_\nu(x)
-\partial_\nu A_\mu(x)
      -ie[A_\mu(x),A_\nu(x)]_*,
\label{eqn:2-4}
\end{eqnarray}
with the Moyal bracket
\beq
[A_\mu(x),A_\nu(x)]_*\equiv
A_\mu(x)*A_\nu(x)-A_\nu(x)*A_\mu(x).
\label{eqn:2-5}
\end{eqnarray}
It determines Lorentz-invariant NC action of the Maxwell sector
\begin{eqnarray} 
{\hat S}_{EM}
&=&-\displaystyle{{1\over 4}}\int\!d^4xd^6\theta W(\theta)
F_{\mu\nu}(x)*F^{\mu\nu}(x).
\label{eqn:2-6}
\end{eqnarray}
\ind
Vertices in Feynman rules in the matter sector are given in 
Fig.~1.\footnote{Maxwell sector will be considered in the end of this 
section.}
Except for kinematical factors they are given by the average
\beq
V(p,q)=\int\!d^6\theta W(\theta) e^{\frac i2p\wedge q}
\equiv \langle e^{\frac i2p\wedge q}\rangle
\label{eqn:2-7}
\end{eqnarray}
with $p\wedge q=p_\mu\theta^{\mu\nu}q_\nu$. It has the properties:
\beq
V(p,p)&=&1\qquad ({\rm normalization}),\nn\\[2mm]
V(q,p)&=&V(p,q)\qquad ({\rm symmetry}),\nn\\[2mm]
V(p',q')&=&V(p,q)\qquad ({\rm Lorentz}\;{\rm invariance}),\nn\\[2mm]
V(p+cq,q)&=&V(p,q)\;\;\;{\rm for}\;{\rm any}\;c
\qquad ({\rm translation}\;{\rm invariance}).
\label{eqn:2-8}
\end{eqnarray}
The normalization is due to the anti-symmetry 
$\theta^{\nu\mu}=-\theta^{\mu\nu}$ and the normalization condition 
(\ref{eqn:1-4}). The symmetry comes from the Lorentz invariance of 
the weight function, $W(-\theta)=W(\theta)$. Lorentz invariance of 
$V(q,p)$ is obvious from the tensor nature of $\theta^{\mu\nu}$. 
The translation invariance (in the momentum space) is also obvious 
from the anti-symmetric nature of $\theta^{\mu\nu}$.
\begin{figure}
\centerline{\includegraphics[width=10.00 cm,height=6.00 cm]
 {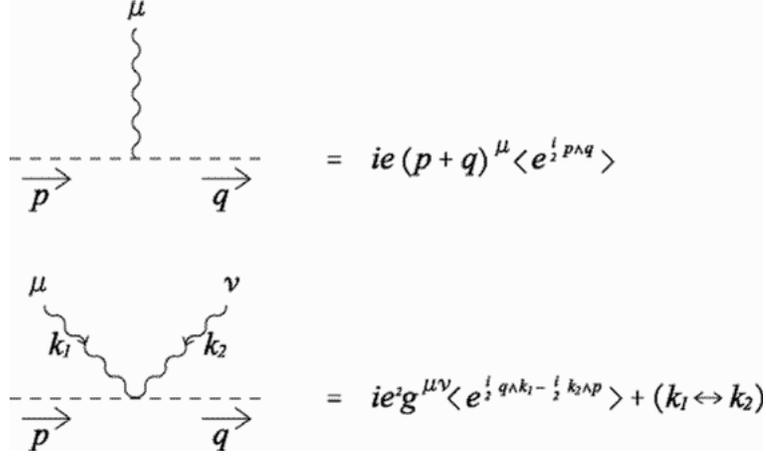}}
 \vspace{1cm}
\caption{Vertices in Lorentz-invariant, scalar NCQED.
Wavy lines for photon and dashed lines for charged scalar.}
\end{figure}
\\
\ind
Photon self-energy diagrams as shown in Fig.~2 sum up to
\begin{figure}
\centerline{\includegraphics[width=12.00 cm,height=5.0 cm]
 {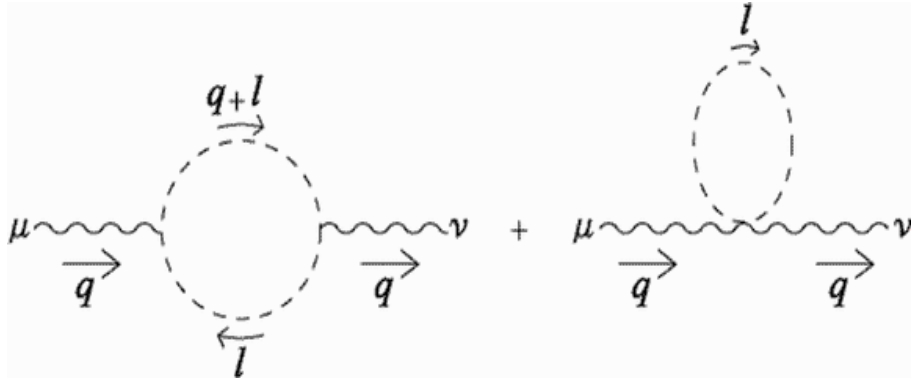}}
   \vspace{1cm}
\caption{Photon self-energy diagrams in scalar QED.}
\end{figure}
\beq
i\Pi_{b(2)}^{\mu\nu}(q)
&=&
e^2\int\!\frac{d^4l}{(2\pi)^4}
\big[
\frac{(2l+q)^\mu(2l+q)^\nu}
{(l^2-m^2+i\epsilon)
((l+q)^2-m^2+i\epsilon)}
\langle e^{\frac i2l\wedge q}\rangle
\langle e^{\frac i2q\wedge l}\rangle\nn\\[2mm]
&&\hspace{1.9cm}-\frac{2g^{\mu\nu}}
{l^2-m^2+i\epsilon}
\langle e^{-iq\wedge l}\rangle\big],
\label{eqn:2-9}
\end{eqnarray}
where $q$ is the momentum of the external photon and $l$ the loop 
momentum. By writing
\beq
\langle e^{-iq\wedge l}\rangle
=\langle e^{\frac i2q\wedge l}\rangle^2
+[\langle e^{-iq\wedge l}\rangle
-\langle e^{\frac i2q\wedge l}\rangle^2]
\label{eqn:2-10}
\eeq
in the second term in (\ref{eqn:2-9}) we can show that the contribution
from the square bracket in (\ref{eqn:2-10}) vanishes in the new UV limit.
(See Appendix B.) We may therefore write using Feynman parameter
\beq
i\Pi_{b(2)}^{\mu\nu}(q)
&=&i\stackrel{\!\!\!(1)}{\Pi_{b(2)}^{\mu\nu}}(q)
+i\stackrel{\!\!\!(2)}{\Pi_{b(2)}^{\mu\nu}}(q),\nn\\[2mm]
i\stackrel{\!\!\!(1)}{\Pi_{b(2)}^{\mu\nu}}(q)
&=&
2e^2\int_0^1\!dx\int\!\frac{d^4l}{(2\pi)^4}
\frac{2l^\mu l^\nu-g^{\mu\nu}(l^2-\Delta)}
{(l^2-\Delta+i\epsilon)^2}
V^2(q,l),\nn\\[2mm]
i\stackrel{\!\!\!(2)}{\Pi_{b(2)}^{\mu\nu}}(q)
&=&
e^2(q^\mu q^\nu-q^2g^{\mu\nu})
\int_0^1\!dx(1-2x)^2\int\!\frac{d^4l}{(2\pi)^4}
\frac {V^2(q,l)}{(l^2-\Delta+i\epsilon)^2},
\label{eqn:2-11}
\end{eqnarray}
where $\Delta=-q^2x(1-x)+m^2$ and we have translated the integration 
variable. Since the extra vertex factor $V^2(q,l)$ depends on 
$q^2, l^2$ as well as $q\cdot l$ by Lorentz invariance, we cannot 
replace $l^\mu l^\nu \to (1/4)g^{\mu\nu}l^2$ in the integrand of 
$\stackrel{\!\!\!(1)}{\Pi_{b(2)}^{\mu\nu}}(q)$ as usually done in 
the symmetric integration. Consequently, we cannot conclude that
the amplitude $\stackrel{\!\!\!(1)}{\Pi_{b(2)}^{\mu\nu}}(q)$ is 
proportional to the metric tensor and, hence, exhibits quadratic 
divergence.\footnote{In other gauge-invariant regularizations 
$\stackrel{\!\!\!(1)}{\Pi_{b(2)}^{\mu\nu}}(q)\to 0$. See the 
next section.}
\\
\ind
To evaluate the integral (\ref{eqn:2-11}) we make Wick 
rotation,\footnote{Wick rotation with respect to $l^0$ is made possible 
in a frame, $q^0\ne 0, \;{\boldsymbol q}={\boldsymbol 0}$. The result is valid for 
generic value of $q$.}
\beq
l^0&=&il_E^4, \quad {\boldsymbol l}={\boldsymbol l}_E,\nn\\[2mm]
q^0&=&iq_E^4, \quad {\boldsymbol q}={\boldsymbol q}_E.
\label{eqn:2-12}
\eeq
Since the theory involves another parameter $\theta^{\mu\nu}$ carrying 
Lorentz indices, we must also perform Wick rotation
\beq
\theta^{0i}\to -i\theta_E^{4i},\;\;\;
\theta^{ij}\to \theta_E^{ij},
\label{eqn:2-13}
\eeq
such that
\beq
p\wedge l=p_E\wedge_El_E\equiv
\sum_{\mu,\nu=1,2,3,4}(p_E)_\mu\theta_E^{\mu\nu}(l_E)_\nu.
\label{eqn:2-14}
\eeq
This is dictated by Lorentz invariance of $V(q,l)$. Then the amplitude 
(\ref{eqn:2-11}) becomes in Euclidean metric with 
$g_E^{\mu\nu}=-\delta^{\mu\nu}$,
\beq
\stackrel{\!\!\!(1)}{\Pi_{b(2)}^{\mu\nu}}(q_E)
&=&
2e^2\int_0^1\!dx\int\!\frac{d^4l_E}{(2\pi)^4}
\frac{2l_E^\mu l_E^\nu+g_E^{\mu\nu}(l_E^2+\Delta_E)}
{(l_E^2+\Delta_E)^2}
V^2(q_E,l_E),\nn\\[2mm]
\stackrel{\!\!\!(2)}{\Pi_{b(2)}^{\mu\nu}}(q)
&=&
e^2(q_E^\mu q_E^\nu+q_E^2g_E^{\mu\nu})
\int_0^1\!dx(1-2x)^2\int\!\frac{d^4l_E}{(2\pi)^4}
\frac {V^2(q_E,l_E)}{(l_E^2+\Delta_E)^2},
\label{eqn:2-15}
\eeq
where $\Delta_E=q_E^2x(1-x)+m^2$. Put
\beq
\int\!\frac{d^4l_E}{(2\pi)^4}
\frac{l_E^\mu l_E^\nu}{(l_E^2+\Delta_E)^2}
V^2(q_E,l_E)&=&C_1g_E^{\mu\nu}
+C_2q_E^\mu q_E^\nu,\nn\\[2mm]
\int\!\frac{d^4l_E}{(2\pi)^4}
\frac 1{l_E^2+\Delta_E}
V^2(q_E,l_E)&=&C_3,
\label{eqn:2-16}
\eeq
where $C_{1,2,3}$ are functions of invariant $q_E^2$. For Gaussian 
weight function which we employ in what follows, they are given by 
(see Appendix A)
\beq
C_1(-q_E^2)&=&-\frac 1{32\pi^2}
\int_0^\infty\!ds 
\frac {\sqrt{s}e^{-s\Delta}}{\big(\sqrt{s+A_Eq_E^2}\big)^5},\nn\\[2mm]
C_2(-q_E^2)&=&
\frac 1{32\pi^2}A_E
\int_0^\infty\!ds 
\frac {e^{-s\Delta}}
{\sqrt{s}\big(\sqrt{s+A_Eq_E^2}\big)^5},\nn\\[2mm]
C_3(-q_E^2)
&=&\frac 1{16\pi^2}
\int_0^\infty\!ds
\frac{e^{-s\Delta}}{\sqrt{s}\big(\sqrt{s+A_Eq_E^2}\big)^3},
\label{eqn:2-17}
\end{eqnarray}
with 
$A_E=\frac {a^4}2\frac{\langle {\bar\theta}_E^{\,2}\rangle}{24}$. It 
follows that
\beq
2C_1(-q_E^2)+C_3(-q_E^2)=2C_2(-q_E^2)q_E^2.
\label{eqn:2-18}
\eeq
Substituting this equation with (\ref{eqn:2-16}) into 
(\ref{eqn:2-15}) yields
\beq
\stackrel{\!\!\!(1)}{\Pi_{b(2)}^{\mu\nu}}(q_E)
&=&
2e^2(q_E^\mu q_E^\nu+q_E^2g_E^{\mu\nu})
\int_0^1\!dx(2C_2(-q_E^2)),\nn\\[2mm]
\stackrel{\!\!\!(2)}{\Pi_{b(2)}^{\mu\nu}}(q)
&=&
e^2(q_E^\mu q_E^\nu+q_E^2g_E^{\mu\nu})
\int_0^1\!dx(1-2x)^2C_4(-q_E^2),
\label{eqn:2-19}
\eeq
where we have defined
\beq
C_4(-q_E^2)=
\int\!\frac{d^4l_E}{(2\pi)^4}
\frac 1{(l_E^2+\Delta_E)^2}
V^2(q_E,l_E)
=\frac 1{16\pi^2}
\int_0^\infty\!ds\frac{\sqrt{s}e^{-s\Delta_E}}
{(\sqrt{s+A_Eq_E^2})^3}.
\label{eqn:2-20}
\eeq
Analytic continuation back to Minkowski metric gives
\beq
\stackrel{\!\!\!(1)}{\Pi_{b(2)}^{\mu\nu}}(q)
&=&
2e^2(q^\mu q^\nu-q^2g^{\mu\nu})
\int_0^1\!dx(2C_2(q^2)),\nn\\[2mm]
\stackrel{\!\!\!(2)}{\Pi_{b(2)}^{\mu\nu}}(q)
&=&
e^2(q^\mu q^\nu-q^2g^{\mu\nu})
\int_0^1\!dx(1-2x)^2C_4(q^2),
\label{eqn:2-21}
\eeq
where $C_i(q^2) (i=1,2,3,4)$ are obtained from $C_i(-q_E^2)$ by 
$q_E^2\to -q^2$ and $A_E\to A=\frac {a^4}2
\frac{\langle {\bar\theta}^{\,2}\rangle}{24}>0$. Thus the piece
$\stackrel{\!\!\!(1)}{\Pi_{b(2)}^{\mu\nu}}$ also becomes transverse as does
$\stackrel{\!\!\!(2)}{\Pi_{b(2)}^{\mu\nu}}$.
This well-come situation is obtained {\it without} cancellation in our 
Lorentz-invariant NCQED. Unfortunately, however, this continuation process
brings about a new problem. The problem arises because, although the 
functions $C_i(-q_E^2)$ are finite for $a^4q_E^2\ne 0$, the functions 
$C_i(q^2)$ are not well-defined for $a^4q^2\ge 0$. This is due to the 
presence of IR singularities in $C_i(-q_E^2)$. \footnote{`Convergent' 
integrals at $a=0$ never possess IR singularity and their analytic 
continuation are defined at $a=0$ so that they are regular in $q^2$ provided
that $\Delta>0$. In 
such case we do not have the identity (\ref{eqn:2-18}) but a different 
one violating the transversality. See the end of the Appendix A.} 
In this respect see, also, the next section.
\\
\ind
To avoid them we may modify $C_i(q^2)$ to cutoff at the lower limit of 
the integration region. As explained in I we instead take the 
regularized functions
\beq
C_2(q^2,\Lambda^2)=\frac 1{32\pi^2}
A\int_0^\infty\!ds 
\frac {e^{-s\Delta-\frac 1{s\Lambda^2}}}
{\sqrt{s}\big(\sqrt{s-Aq^2}\big)^5},\nn\\[2mm]
C_4(q^2,\Lambda^2)=\frac 1{16\pi^2}
\int_0^\infty\!ds 
\frac {\sqrt{s}e^{-s\Delta-\frac 1{s\Lambda^2}}}
{\big(\sqrt{s-Aq^2}\big)^3}.
\label{eqn:2-22}
\end{eqnarray}
The cutoff factor $e^{-\frac 1{s\Lambda^2}}$ is introduced to avoid 
the singularity at $s=Aq^2$ by taking the new UV limit (\ref{eqn:1-7})
and the parameter $\Lambda$ is qualified to be called UV cutoff 
because it effectively cuts off the lower limit of Schwinger's 
$s$-integration. Since $C_2$ goes like $A\Lambda^4/(32\pi^2)$ in the 
new UV limit, we impose the condition
\beq
\Lambda^2a^2\to 0
\label{eqn:2-23}
\end{eqnarray}
to eliminate it since $A$ is proportional to $a^4$.\footnote{This
is mentioned only in a footnote of I.} In other words, 
the first term in (\ref{eqn:2-11})
vanishes in the new UV limit supplemented with (\ref{eqn:2-23}) as 
in other regularizations. The new UV limit of $C_4$ turns out to be 
given by
\beq
\mathop{\rm lim}_{\Lambda^2\to\infty,
a^2\to 0, \Lambda^2a^2:{\rm fixed}}\;
C_4(q^2,\Lambda^2)
&=&\mathop{\rm lim}_{\Lambda^2\to\infty}\,
\frac 1{8\pi^2}
K_0\big(2\sqrt{\Delta/\Lambda^2}\big),
\label{eqn:2-24}
\eeq
where $K_0$ is the modified Bessel function of the second kind. 
Subtraction at $q^2=0$ leads to
\beq
\Pi_{b(2),R}^{\mu\nu}(q)
&=&\stackrel{\!\!\!(2)}{\Pi_{b(2),R}^{\mu\nu}}(q)=
\frac \alpha{4\pi}(q^\mu q^\nu-q^2g^{\mu\nu})
\int_0^1\!dx(1-2x)^2\ln{\big(\frac{m^2}
{\Delta}\big)}.
\label{eqn:2-25}
\end{eqnarray}
This is the renormalized photon self-energy amplitude obtained 
through NC regularization and the same as obtained in other gauge-invariant
regularizations.
\\
\ind
Let us now consider the Maxwell sector. In order to consistently quantize 
the gauge field in NCQED it is necessary to introduce the ghost fields, 
$c,{\bar c}$, and the Nakanishi-Lautrup field $B$ such that the full action 
is BRST-invariant.\cite{14),9)} We use the Feynman rules of
Ref.~9) as shown in Fig.~3 and choose the Feynman-'t Hooft gauge.
\begin{figure}
\centerline{\includegraphics[width=12.00 cm,height=10.00 cm]
 {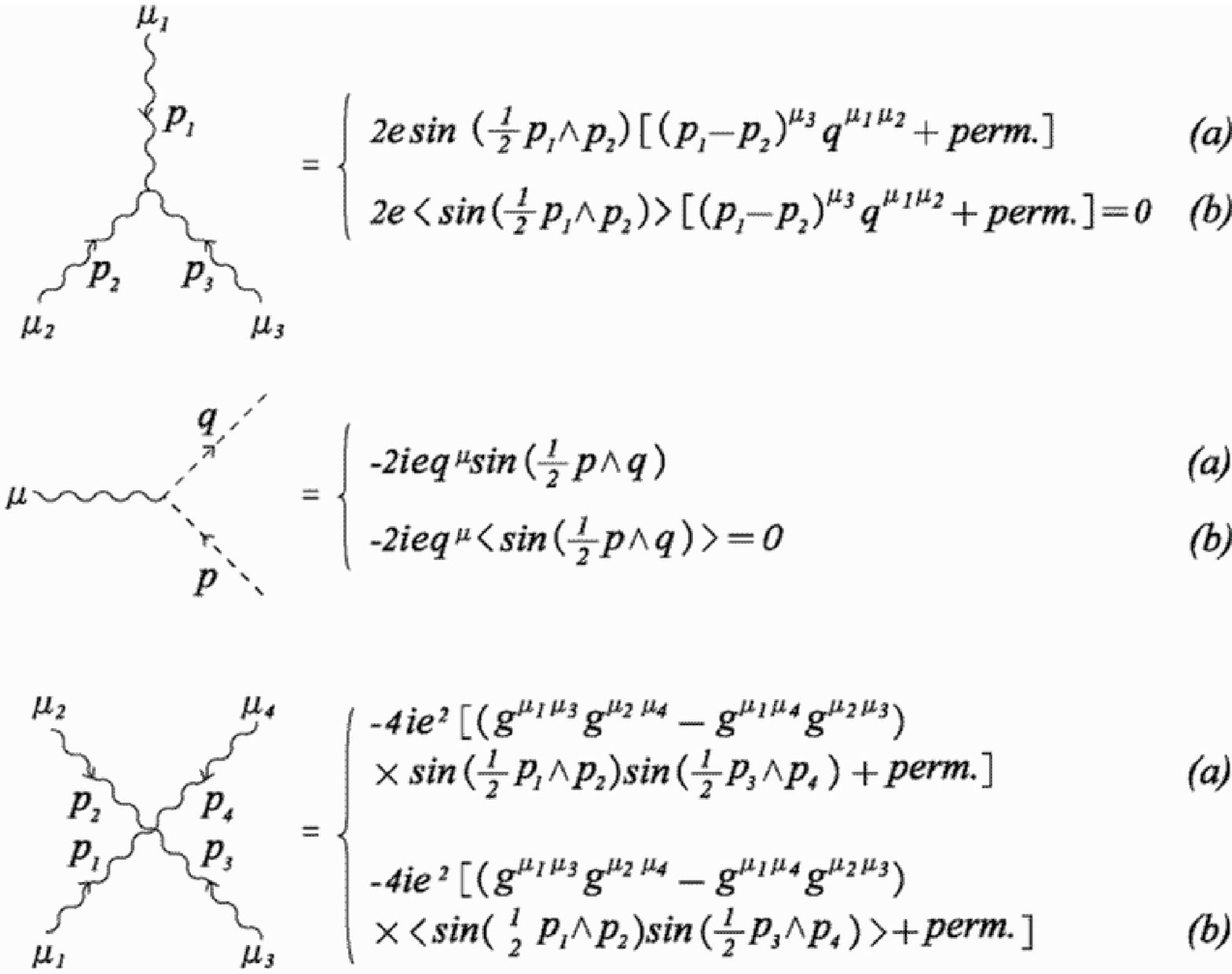}}
\caption{Feynman rules in the Maxwell sector of NCQED (a) and
Lorentz-invariant NCQED (b). Wavy line represents photon and
dashed line with arrow ghost.}
\end{figure}
Note that there exist no three-point vertices in the Lorentz-invariant 
NCQED if the action (\ref{eqn:2-6}) is employed, because 
$\langle\sin{(\frac 12p\wedge q)}\rangle=0$. Consequently, ghosts 
decouple and there is only one more contribution to the 
photon self energy, the tadpole diagram. \\
\ind
The tadpole diagram as shown in Fig.~4 is given by
\begin{figure}
\centerline{\includegraphics[width=13.00 cm,height=5.500 cm]
 {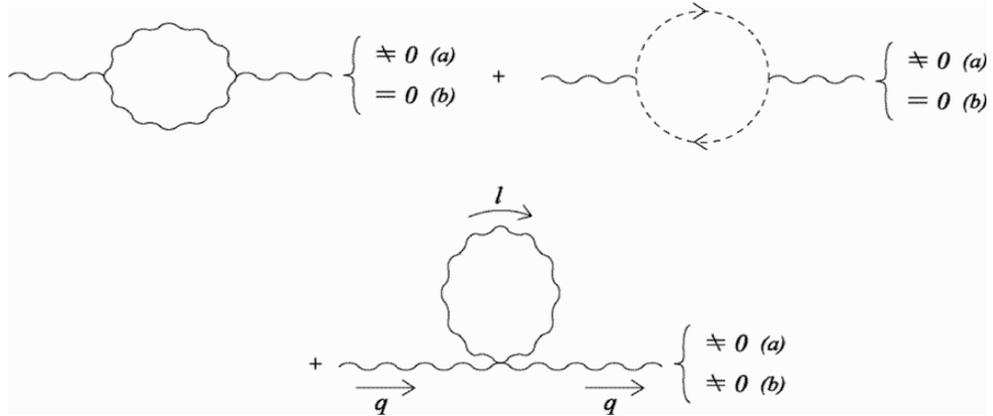}}
\vspace{1cm}
\caption{Photon self-energy diagrams in the Maxwell sector of NCQED. 
All diagrams contribute in NCQED (a), while only tadpole diagram 
has to be considered in Lorentz-invariant NCQED (b).}
\end{figure}
\beq
i\Pi_{\rm tadpole(2)}^{\mu\nu}(q)
&=&-12e^2g^{\mu\nu}
\int\!\frac{d^4l}{(2\pi)^4}
\frac 1{l^2+i\epsilon}\langle\sin^2{(\frac 12q\wedge l)}\rangle\nn\\[2mm]
&=&-6e^2g^{\mu\nu}
\big[\int\!\frac{d^4l}{(2\pi)^4}
\frac 1{l^2+i\epsilon}-\int\!\frac{d^4l}{(2\pi)^4}
\frac 1{l^2+i\epsilon}\langle\cos{(q\wedge l)}\rangle\big],
\label{eqn:2-26}
\eeq
where $q$ denotes the external photon momentum. As shown in I the new 
UV limit of the tadpole diagram is proportional to $a^4\Lambda^4$ and 
vanishes if we impose the condition (\ref{eqn:2-23}). All contributions 
arising from non-Abelian nature of Lorentz-invariant NC Maxwell action 
(\ref{eqn:2-6}) with ghost and gauge-fixing terms included disappear 
at least at one-loop order. In conclusion NC regularization with 
(\ref{eqn:2-23}) gives rise to the renormalized one-loop photon 
self-energy in scalar QED given by (\ref{eqn:2-25}) in accordance 
with other regularization schemes.
%%%%%%%%%%%%%%%%%%%%%%%%%%%%%%%%%%%%%%%%%%%%%%%%%%%%%%%%%%
\section{Vacuum polarization in QED} 
%%%%%%%%%%%%%%%%%%%%%%%%%%%%%%%%%%%%%%%%%%%%%%%%%%%%%%%%%%
This section is devoted to a compact presentation of our previous\cite{1)}
treatment of vacuum polarization in NC regularization scheme
comparing with Pauli-Villars-Gupta and dimensional regularizations.
The basic mathematical formulae we needed in I
have repeatedly been used in the previous
section and are collected in the Appendix A.
\\
\ind
According to (\ref{eqn:1-2}) the matter sector of the 
Lorentz-invariant NCQED is defined by the action
\beq
\hS_{\rm D}
&=&
\int\!d^4xd^6\theta W(\theta)\big[{\bar\psi}(x)
(i\gamma^\mu\partial_\mu-M)\psi(x)+
e{\bar\psi}(x)*
\gamma^\mu A_\mu(x)*\psi(x)\big].
\label{eqn:3-1}
\eeq
The spinor is subject to the $\ast$-gauge transformation as in 
(\ref{eqn:2-2}) and has the same covariant derivative,
\beq
\psi(x)&\to& ^{\hat g}\psi(x)
=U(x)*\psi(x),\;\;\;
U(x)*U^{\dag}(x)=U^{\dag}(x)*U(x)=1,\nn\\[2mm]
D_\mu\psi(x)&=&\partial_\mu\psi(x)-ieA_\mu(x)*\psi(x).
\label{eqn:3-2}
\eeq
The gauge field $A_\mu$ transforms as in (\ref{eqn:2-3}).
The discrete symmetries of Lorentz-invariant NCQED
can be shown as in Ref.~16) in which we assumed
fields to depend on $x$ as well as $\theta$.
What we need in the present case is to delete the additional
`dependence' of fields on $\theta$.
\\
\ind
Using the action (\ref{eqn:3-1}) the vacuum polarization tensor in 
Lorentz-invariant NCQED is given by (see Fig.~5)
\begin{figure}
\centerline{\includegraphics[width=8.5 cm,height=6.00 cm]
 {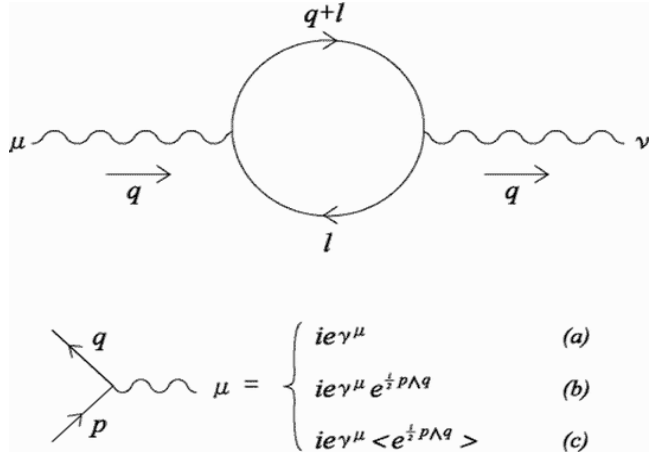}}
 \vs
 \vs
 \vs
 \caption{Vacuum polarization. Wavy lines for photon and
 solid lines for fermion. Vertices in QED (a), NCQED (b) and
 Lorentz-invariant NCQED (c).}
\end{figure}

\beq
i\Pi_{f(2)}^{\mu\nu}(q)&=&
(ie)^2(-1)\int\!\frac{d^4l}{(2\pi)^4}
{\rm Tr}\big[
\gamma^\mu\frac i{\lslash-M+i\epsilon}
\gamma^\nu\frac i{\qslash+\lslash-M+i\epsilon}\big]%\nn\\[2mm]
%&&\qquad\qquad\qquad\times
\langle e^{\frac i2q\wedge l}\rangle
\langle e^{\frac i2l\wedge q}\rangle,
\label{eqn:3-3}
\eeq
where $q$ is the external photon momentum and $l$ the loop momentum.
The vacuum polarization tensor in QED and NCQED\footnote{Vacuum
polarization in NCQED is obtained by replacing the average
$\langle e^{\frac i2q\wedge l}\rangle$ with the Moyal
phase $e^{\frac i2q\wedge l}$. Two Moyal phases in (\ref{eqn:3-3}) 
without the average brackets cancel out and the result is the same
as in QED.} is given by (\ref{eqn:3-3})
without the extra vertex factors. It is denoted 
$\Pi_{f(2)}^{\mu\nu}(q)_0$ below. Computing the Dirac trace and 
translating the integration variable we have a similar expression
like the scalar case,
\beq
\Pi_{f(2)}^{\mu\nu}(q)
&=&\stackrel{\!\!\!(1)}{\Pi_{f(2)}^{\mu\nu}}(q)
+\stackrel{\!\!\!(2)}{\Pi_{f(2)}^{\mu\nu}}(q),\nn\\[2mm]
i\stackrel{\!\!\!(1)}{\Pi_{f(2)}^{\mu\nu}}(q)
&=&
-4e^2\int_0^1\!dx\int\!\frac{d^4l}{(2\pi)^4}
\frac{2l^\mu l^\nu-g^{\mu\nu}(l^2-\Delta)}
{(l^2-\Delta+i\epsilon)^2}
V^2(q,l),\nn\\[2mm]
i\stackrel{\!\!\!(2)}{\Pi_{f(2)}^{\mu\nu}}(q)
&=&
4e^2(q^\mu q^\nu-q^2g^{\mu\nu})
\int_0^1\!dx2x(1-x)\int\!\frac{d^4l}{(2\pi)^4}
\frac {V^2(q,l)}{(l^2-\Delta+i\epsilon)^2},
\label{eqn:3-4}
\end{eqnarray}
with $\Delta=-q^2x(1-x)+M^2$. Before computing `non-transverse' part
$\stackrel{\!\!\!(1)}{\Pi_{f(2)}^{\mu\nu}}\!\!(q)$
in NC regularization, let us first consider it in Pauli-Villars-Gupta
and dimensional regularizations when no extra vertex factors appear
as in QED and NCQED. Put
\beq
I^{\mu\nu}(q,M)&=&-i
\int\!\frac{d^4l}{(2\pi)^4}
\frac{2l^\mu l^\nu-g^{\mu\nu}(l^2-\Delta)}
{(l^2-\Delta+i\epsilon)^2}\nn\\[2mm]
&=&\int\!\frac{d^4l_E}{(2\pi)^4}
\frac{2l_E^\mu l_E^\nu-g_E^{\mu\nu}(-l_E^2-\Delta)}
{(l_E^2+\Delta)^2},
\label{eqn:3-5}
\eeq
where we have made Wick rotation. It is integrated over $x$ from 0 to 1 
to give $\stackrel{\!\!\!(1)}{\Pi_{f(2)}^{\mu\nu}}\!\!(q)_0$ apart from a 
constant. By symmetric integration
$l^\mu l^\nu\to (1/4)g^{\mu\nu}l^2$ in the first integrand (or
$l_E^\mu l_E^\nu\to -(1/4)g_E^{\mu\nu}l_E^2$ in the second integrand).
Using Schwinger representation we obtain
\beq
I^{\mu\nu}(q,M)&=&g_E^{\mu\nu}
\int_0^\infty\!dss\int\!\frac{d^4l}{(2\pi)^4}
(\frac 12l_E^2+\Delta)e^{-s(l_E^2+\Delta)}
\nn\\[2mm]
&=&-g_E^{\mu\nu}
\frac 1{16\pi^2}
\int_0^\infty\!ds\frac\partial{\partial s}
\big(\frac 1se^{-s\Delta)}\big)
=
-g_E^{\mu\nu}
\frac 1{16\pi^2}
\frac 1se^{-s\Delta}|_{s=0}^{s=\infty}.
\label{eqn:3-6}
\eeq
The lower limit does not exist (provided $\Delta$ is assumed to be
positive). Pauli-Villars-Gupta regularization to cure 
this defect consists of replacing the integral $I^{\mu\nu}(q,M)$ with
$I^{\mu\nu}_{reg}(q)=\sum_{i=0,1,2}C_iI^{\mu\nu}(q,M_i)$
such that $\sum_{i=0,1,2}C_i=0$ and $\sum_{i=0,1,2}C_iM_i^2=0$,
where $C_0=1, M_0=M$. It follows that $I^{\mu\nu}_{reg}(q)=0$ because
\beq
I^{\mu\nu}_{reg}(q)
&=&
g_E^{\mu\nu}
\frac 1{16\pi^2}\sum_{i=0,1,2}C_i
\big(
\frac 1se^{-s\Delta_i}\big)|_{s=0}\nn\\[2mm]
&=&
g_E^{\mu\nu}
\frac 1{16\pi^2}\sum_{i=0,1,2}C_i
\big(\frac 1s+q^2x(1-x)-M_i^2\big)|_{s=0}=0.
\label{eqn:3-7}
\eeq
On the other hand, dimensional regularization extends dimensions 4$\to n$ 
in which case $l^\mu l^\nu\to(1/n)g^{\mu\nu}l^2$ by symmetric integration.
Then we have again using Schwinger representation
\beq
I_{4\to n}^{\mu\nu}(q,M)
&=&g_E^{\mu\nu}
\int\!\frac{d^nl_E}{(2\pi)^n}
\frac{(1-\frac 2n)l_E^2+\Delta}
{(l_E^2+\Delta)^2}\nn\\[2mm]
&=&
g_E^{\mu\nu}\frac 1{\Gamma(n/2)(4\pi)^{n/2}}\nn\\[2mm]
&&\hspace{0.5cm}\times\int_0^\infty\!ds
\big((1-2/n)\Gamma(n/2+1)
s^{-n/2}
+\Delta\Gamma(n/2)s^{-n/2+1}\big)
e^{-s\Delta}
=0,
\label{eqn:3-8}
\eeq
where we have used $\Gamma(z+1)=z\Gamma(z)$.
In either case we regularize 
$\stackrel{\!\!\!(1)}{\Pi_{f(2)}^{\mu\nu}}\!\!(q)_0$ to zero
satisfying gauge invariance. 
\\
\ind
On the contrary, NC regularization `dispenses', in a sense, with the 
above regularization.
We directly integrates the amplitudes (\ref{eqn:3-4}) for Gaussian weight 
function, which turn out to be {\it finite} for $a^4q^2<0$. The procedure is 
already illustrated in scalar QED and detailed in I. By Wick rotation 
(\ref{eqn:2-12}) through (\ref{eqn:2-14}) we obtain (see (\ref{eqn:2-21}))
\beq
\stackrel{\!\!\!(1)}{\Pi_{f(2)}^{\mu\nu}}\!\!(q)
&=&
-4e^2(q^\mu q^\nu-q^2g^{\mu\nu})
\int_0^1\!dx(2C_2(q^2)),\nn\\[2mm]
\stackrel{\!\!\!(2)}{\Pi_{f(2)}^{\mu\nu}}\!\!(q)
&=&
4e^2(q^\mu q^\nu-q^2g^{\mu\nu})
\int_0^1\!dx2x(1-x)C_4(q^2).
\label{eqn:3-9}
\end{eqnarray}
Since $C_{2,4}(q^2)$ are finite for $a^4q^2<0$, (\ref{eqn:3-9}) give 
{\it finite}, transverse vacuum polarization tensor in Lorentz-invariant 
NCQED (\ref{eqn:3-1}) in the same region. At first sight this conclusion 
seems to differ from that of the known regularizations, 
$\stackrel{\!\!\!(1)}{\Pi_{f(2)}^{\mu\nu}}\!\!(q)_0\to 0$, in QED.
However, it is possible to fill up this apparent difference
by noting that the commutative limit $a\to 0$ cannot be
interchangeable with UV limit $\Lambda\to\infty$, that is, they must be taken 
{\it simultaneously} according to the new UV limit (\ref{eqn:1-7}) {\it with}
the condition (\ref{eqn:2-23}). To be more precise we replace $C_{2,4}(q^2)$ 
by the regularized functions (\ref{eqn:2-22}) and take the new UV limit
(\ref{eqn:1-7}) with (\ref{eqn:2-23}). It can then be shown in exactly the 
same way as in scalar QED that the piece 
$\stackrel{\!\!\!(1)}{\Pi_{f(2)}^{\mu\nu}}\!\!(q)$ vanishes, leaving the 
well-known result after subtraction at $q^2=0$,
\beq
\Pi_{f(2),\lower 1.2pt \hbox{{\scriptsize ${R}$}}}^{\mu\nu}(q)
&=&(q^\mu q^\nu-q^2g^{\mu\nu})
\big(-\frac{2\alpha}\pi\big)
\int_0^1\!dxx(1-x)\ln{\big(\frac{M^2}\Delta\big)}.
\label{eqn:3-10}
\eeq
The Maxwell sector is the same as in scalar QED and need not be repeated 
here. (See also I.)
%%%%%%%%%%%%%%%%%%%%%%%%%%%%%%%%%%%%%%%%%%%%%%%%%%%%%%%%%%
\section{One-loop gluon self-energy in $U(N)$ gauge theory} 
%%%%%%%%%%%%%%%%%%%%%%%%%%%%%%%%%%%%%%%%%%%%%%%%%%%%%%%%%%
In this section we consider $U(N)$ gauge theory without matter. The basic 
field variable in the theory is $U(N)$ gauge field 
$A_\mu(x)=\sum_{A=0\cdots N^2-1} T_AA_\mu^A(x)$, where 
$T_A, A=0,1,\cdots,N^2-1$ denote $U(N)$ generators with
\beq
[T_A,T_B]&=&if_{ABC}T_C,\nn\\[2mm]
\{T_A,T_B\}&=&d_{ABC}T_C,\nn\\[2mm]
{\rm Tr}T_AT_B&=&\frac 12\delta_{AB}. 
\label{eqn:4-1}
\eeq
Following Ref.~15) we label $SU(N)$ components  
by small letters, say, $a=1,2,\cdots, N^2-1$. The structure constant
$f_{ABC}$ equals $f_{abc}$ for $SU(N)$ components and $f_{0BC}=f_{A0C}=0$.
The gauge field transforms as
\beq
A_\mu(x)&\to&^gA_\mu(x)=
U(x)*A_\mu(x)*U^{\dag}(x)+\frac igU(x)*\partial_\mu U^{\dag}(x),\nn\\[2mm]
&&U(x)*U^{\dag}(x)=U^{\dag}(x)*U(x)=1.
\label{eqn:4-2}
\eeq
This gauge transformation mixes $U(1)$ component $A_\mu^0$ with $SU(N)$ 
ones $A_\mu^a$. NC non-Abelian gauge field strength takes of the same form 
as that in (\ref{eqn:2-4}),
\begin{eqnarray} 
F_{\mu\nu}(x)&=&\partial_\mu A_\nu(x)
-\partial_\nu A_\mu(x)
      -ig[A_\mu(x),A_\nu(x)]_*,
\label{eqn:4-3}
\end{eqnarray}
where the nonlinear term is decomposed as follows.
\beq
[A_\mu(x),A_\nu(x)]_*&\equiv&
A_\mu(x)*A_\nu(x)-A_\nu(x)*A_\mu(x)\nn\\[2mm]
&=&\frac 12\sum_{A,B=0,1,\cdots,N^2-1}
(A_\mu^A(x)*A_\nu^B(x)-A_\nu^B(x)*A_\mu^A(x))
\{T_A,T_B\}\nn\\[2mm]
&&+\frac 12\sum_{a,b=1,\cdots,N^2-1}
(A_\mu^a(x)*A_\nu^b(x)+A_\nu^b(x)*A_\mu^a(x))
[T_a,T_b].
\label{eqn:4-4}
\end{eqnarray}
To obtain the last term we used the relation $f_{0BC}=f_{A0C}=0$.
Consequently, only in the Moyal bracket term appears the zeroth component.
The fact that (\ref{eqn:4-4}) contains not only the commutators but also 
the anti-commutators of generators explicitly demonstrates that $U(1)$ 
is not decoupled from $SU(N)$ in the field strength. 
\\
\ind
Lorentz-invariant, gauge-fixed action of NC $U(N)$ YM is given by
\begin{eqnarray} 
{\hat S}
&=&\int\!d^4xd^6\theta W(\theta){\rm Tr}
\big[
-\displaystyle{{1\over 4}}F_{\mu\nu}(x)*F^{\mu\nu}(x)
-\frac 1{2\xi}(\partial_\mu A^\mu)^2
+\frac 12
(i{\bar c}*\partial_\mu D^\mu c
-\partial_\mu D^\mu c*{\bar c})\big],
\label{eqn:4-5}
\end{eqnarray}
where covariant derivative of the ghost field is given by
\beq
D_\mu c=\partial_\mu c-ig[A_\mu, c]_*.
\label{eqn:4-6}
\end{eqnarray}
Our prescription leading to the above action is based on (\ref{eqn:1-2}) 
using gauge-fixed action of NC $U(N)$ YM in Ref.16). We employ 
Feynman-'t Hooft gauge $\xi=1$ as in \S 2. Feynman rules for NC $U(N)$ 
gauge theory without $\theta$-integration are given in Refs.~15) and 17).
We need Feynman rules derived from (\ref{eqn:4-5}). They are simply given by
integrating those in Refs.~15) and 17) over $\theta$ at each vertex.
The result is displayed in Fig.~6. 
\begin{figure}
\centerline{\includegraphics[width=13.00 cm,height=10.00 cm]
 {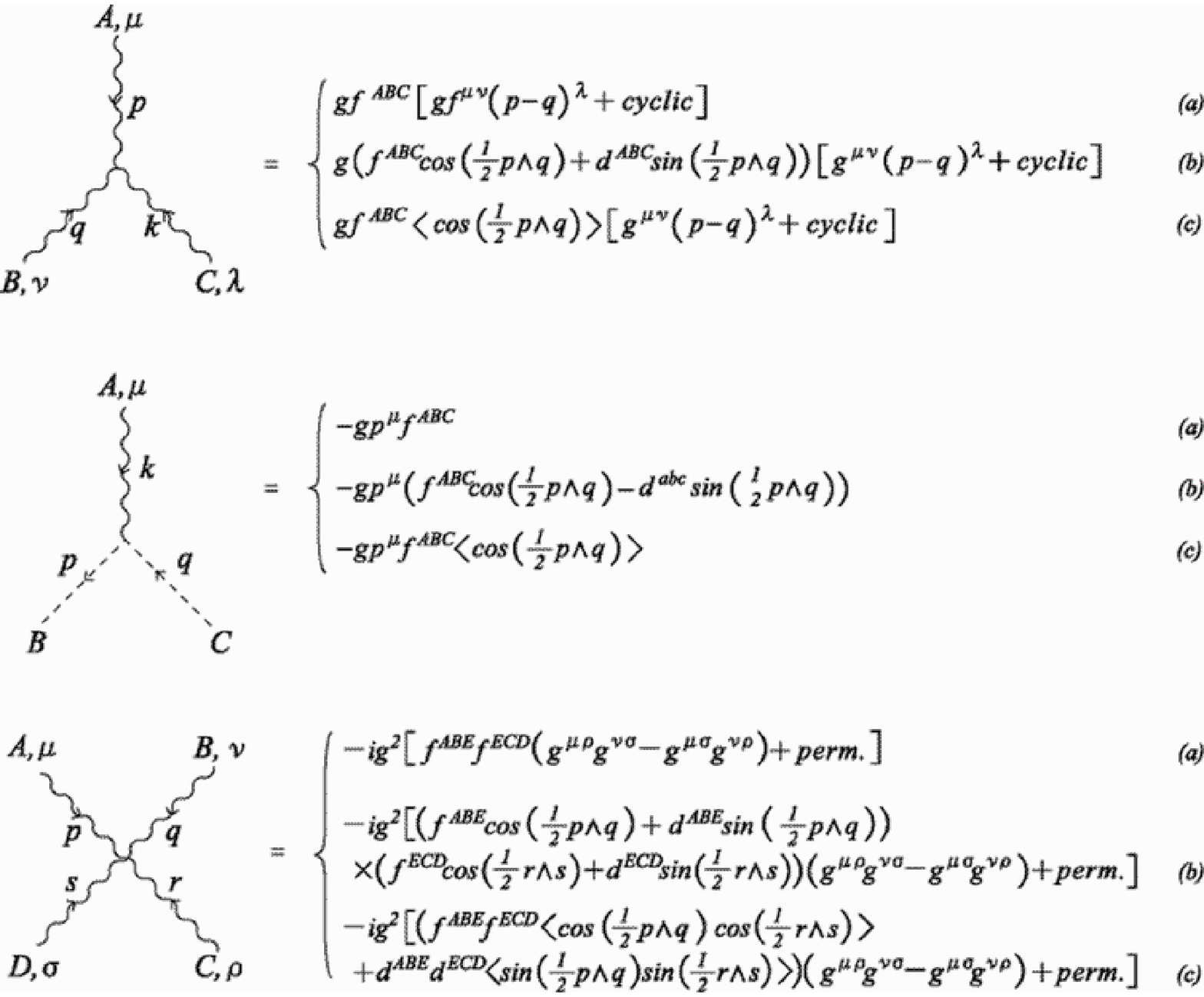}}
\vspace{1cm}
\caption{Feynman rules in $U(N)$ Yang-Mills (a),
NC $U(N)$ Yang-Mills (b) and Lorentz-invariant $U(N)$ Yang-Mills.
Wavy lines for gauge bosons and dashed lines with arrow for ghosts.}
\end{figure}
It is seen that $\theta$-integration
helps decouple $U(1)$ from $SU(N)$ in all three-point vertices. This 
implies, in particular, that the zeroth component of ghost field completely
decouples from the theory. On the other hand, $U(1)$ gauge boson couples 
to $SU(N)$ gauge boson only through 4-point vertex with the extra vertex 
factor carrying $\langle \sin{\frac 12(p\wedge q)}
\sin{\frac 12(r\wedge s)}\rangle$ where $p,q,r,s$ are momenta flowing 
into the vertex.
\\
\ind
One-loop self-energy correction of $SU(N)$ gauge boson is given by the 
sum of diagrams as shown in Fig.~7. 
\begin{figure}
\centerline{\includegraphics[width=11.0 cm,height=7.00 cm]
 {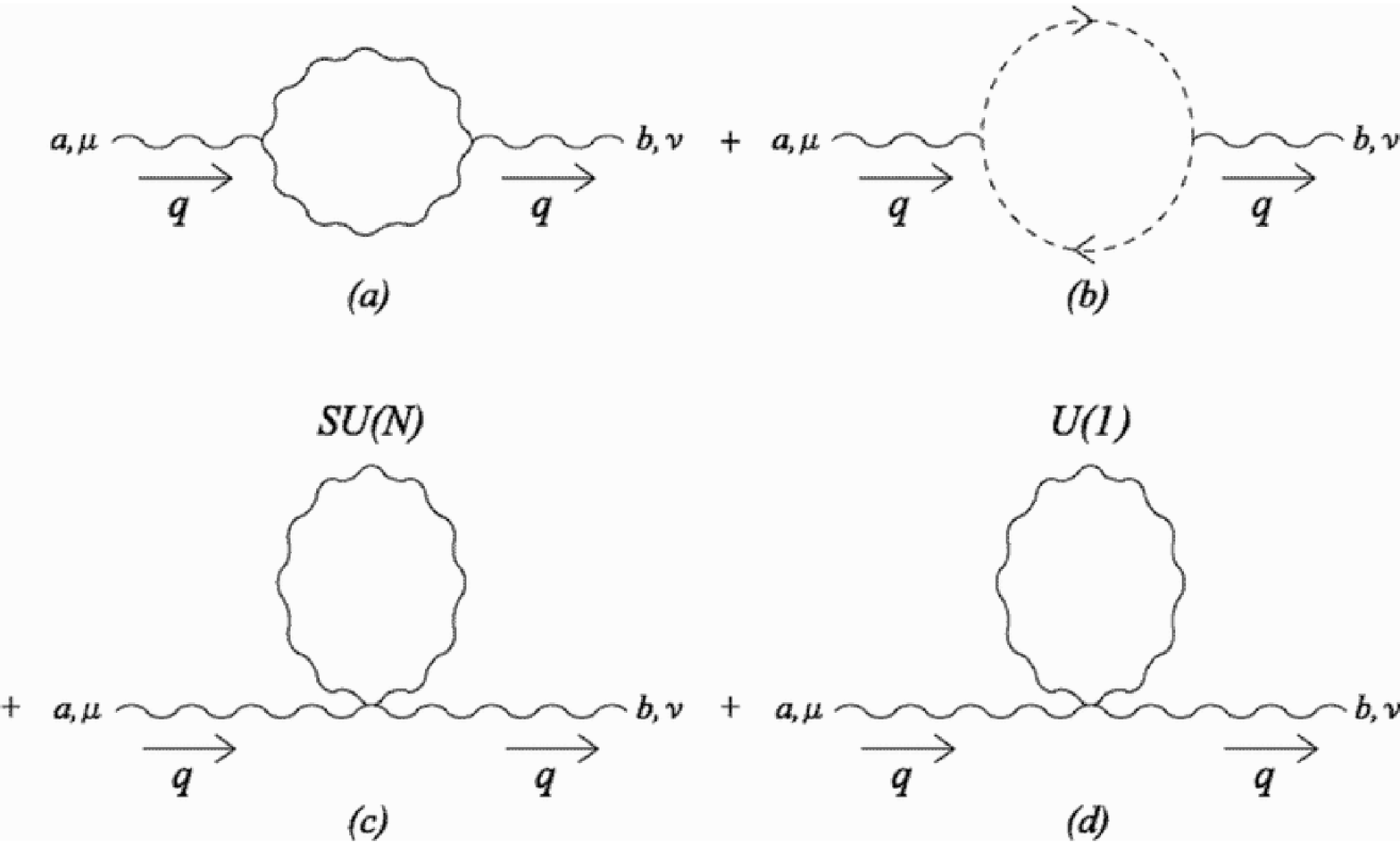}}
 \vspace{1cm}
\caption{One-loop gluon self-energy diagrams. (a), (b), (c) contain
only $SU(N)$ loops, while  $U(1)$  circulates in the loop (d)
in Lorentz-invariant NC $U(N)$ YM.}
\vspace{0.5cm}
\end{figure}
Ignoring the last diagram Fig.~7(d)
for the moment and replacing the extra vertex factor 
$\langle \cos^2{\frac 12(l\wedge q)}\rangle$ in the tadpole diagram
with $\langle \cos{\frac 12(l\wedge q)}\rangle^2$,\footnote{The difference
contributes nothing in the new UV limit as in QED and we present 
an explicit proof in the Appendix C. In what follows we make use of this 
replacement in all tadpole diagrams.} we find the following result in 
terms of the invariant functions defined by (\ref{eqn:2-17}) and 
(\ref{eqn:2-20}):
\beq
{\rm Fig.~7(a)}&=&
\frac 12g^2f_{acd}f_{bcd}\int\!d^4l\frac {-i}{l^2+i\epsilon}
\frac {-i}{(l+q)^2+i\epsilon}
[g^{\mu\rho}(q-l)^\sigma+g^{\rho\sigma}(q+2l)^\mu
+g^{\sigma\mu}(-2q-l)^\rho]\nn\\[2mm]
&&\times
[\delta^\nu_{\;\;\rho}(l-q)_\sigma
+g_{\rho\sigma}(-q-2l)^\nu
+\delta_\sigma^{\;\;\nu}(2q+l)_\rho]
\langle \cos{\frac 12(l\wedge q)}\rangle^2
\nn\\[2mm]
&=&-\frac i2g^2C_2(SU(N))\delta_{ab}
\int_0^1\!dx\big[
g_E^{\mu\nu}\{2C_3(-q_E^2)
+q_E^2(4x^2-10x+8)C_4(-q_E^2)
-10C_1(-q_E^2)\}\nn\\[2mm]
&&\hspace{4.8cm}+q_E^\mu q_E^\nu \{
-10 C_2(-q_E^2)
-(10x^2-10x-2)C_4(-q_E^2)\}\big]\nn\\[2mm]
{\rm Fig.~7(b)}&=&
(-1)g^2f_{acd}f_{bdc}\int\!d^4l\frac {i}{l^2+i\epsilon}
\frac {i}{(l+q)^2+i\epsilon}
(l+q)^\mu l^\nu\langle \cos{\frac 12(l\wedge q)}\rangle^2
\nn\\[2mm]
&=&-ig^2C_2(SU(N))\delta_{ab}
\int_0^1\!dx\big[
g_E^{\mu\nu}C_1(-q_E^2)
+q_E^\mu q_E^\nu(C_2(-q_E^2)
-x(1-x)C_4(-q_E^2))\big],\nn\\[2mm]
{\rm Fig.~7(c)}&=&
\frac 12(-ig)^2f_{acd}f_{bdc}\int\!d^4l\frac {-i}{l^2+i\epsilon}
6g^{\mu\nu}\langle \cos{\frac 12(l\wedge q)}\rangle^2
\nn\\[2mm]
&=&
ig^2C_2(SU(N))\delta_{ab}3g_E^{\mu\nu}
\int_0^1\!dx\big[
C_3(-q_E^2)
+q_E^2(2x^2-3x+1)C_4(-q_E^2)\big],
\label{eqn:4-7}
\end{eqnarray}
where $C_2(SU(N))$ is the second Casimir. Using the identity (\ref{eqn:2-18})
to eliminate $C_{1,3}$ in favor of $C_2$, we obtain
\beq
i\Pi_{g(2)}^{\mu\nu,ab}(q_E)&\equiv&
{\rm Fig.~7(a)}+{\rm Fig.~7(b)}+{\rm Fig.~7(c)}
=
ig^2C_2(SU(N))\delta_{ab}
(q_E^\mu q_E^\nu+q_E^2g_E^{\mu\nu})
\pi_{(2)}(-q_E^2),\nn\\[2mm]
\pi_{(2)}(-q_E^2)&=&
\int_0^1\!dx
[4C_2(-q_E^2)+(4x^2-4x-1)C_4(-q_E^2)].
\label{eqn:4-8}
\end{eqnarray}
Analytic continuation back to Minkowski metric finally gives
\beq
\Pi_{g(2)}^{\mu\nu,ab}(q)
&=&
g^2C_2(SU(N))\delta_{ab}
(q^\mu q^\nu-q^2g^{\mu\nu})
\pi_{(2)}(q^2),\nn\\[2mm]
\pi_{(2)}(q^2)&=&
\int_0^1\!dx
[4C_2(q^2)+(4x^2-4x-1)C_4(q^2)].
\label{eqn:4-9}
\end{eqnarray}
Consequently, the amplitude $\Pi_{g(2)}^{\mu\nu,ab}(q)$, 
which is finite for $a^4q^2<0$, becomes transverse as in QED. As we 
have seen in \S 2, IR singularity in $C_{2,4}(-q_E^2)$ is eliminated by 
employing the regularized functions (\ref{eqn:2-20}) which lead, in 
the new UV limit under the condition (\ref{eqn:2-23}), to
$C_2\to 0$ and $C_4\to(1/8\pi^2)K_0(2\sqrt{\Delta/\Lambda^2})$
(see (\ref{eqn:2-24})). Hence, $\pi_{(2)}(q^2)$ exhibits
log divergence as should be the case.
Subtraction at $q^2=-\mu^2$ yields
\beq
\Pi_{g(2),R}^{\mu\nu,ab}(q)
&=&
\frac{g^2}{4\pi}C_2(SU(N))\delta_{ab}
(q^\mu q^\nu-q^2g^{\mu\nu})
\pi_{(2),R}(q^2),\nn\\[2mm]
\pi_{(2),R}(q^2)&=&
\int_0^1\!dx(4x^2-4x-1)\ln{\big(\frac{\mu^2}{-q^2}\big)}
=-\frac 53\ln{\big(\frac{\mu^2}{-q^2}\big)}.
\label{eqn:4-10}
\end{eqnarray}
\ind
The diagram as shown in Fig.~7(d), in which $U(1)$ gauge boson circulates
in the loop, gives \footnote{In what follows we use the complete symmetry
of $d_{ABC}$ with $d_{0ab}=\sqrt{\frac 2N}\delta_{ab}, d_{00a}=0$,
and $d_{000}=\sqrt{\frac 2N}$.}
\beq
{\rm Fig.~7(d)}&=&
-3g^2\frac 2N\delta_{ab}g^{\mu\nu}
\int\!\frac{d^4l}{(2\pi)^4}
\frac 1{l^2+i\epsilon}\langle \sin^2{\frac 12(q\wedge l)}
\rangle.
\label{eqn:4-11}
\end{eqnarray}
This is essentially equal to (\ref{eqn:2-26}). As noted there the 
new UV limit of (\ref{eqn:4-11}) vanishes upon using the condition 
(\ref{eqn:2-23}). To sum up one-loop self-energy amplitude of $SU(N)$ 
gauge boson is given by (\ref{eqn:4-10}) which include only $SU(N)$ 
gauge bosons circulating in the loop.
\\
\ind
As for $U(1)$ gauge boson one-loop self-energy diagrams are given 
by Figs.~8(a) and 8(b) where $SU(N)$ loop and $U(1)$ loop are 
considered, respectively;
\vspace{1cm}
\begin{figure}
\centerline{\includegraphics[width=11.00 cm,height=3.5 cm]
 {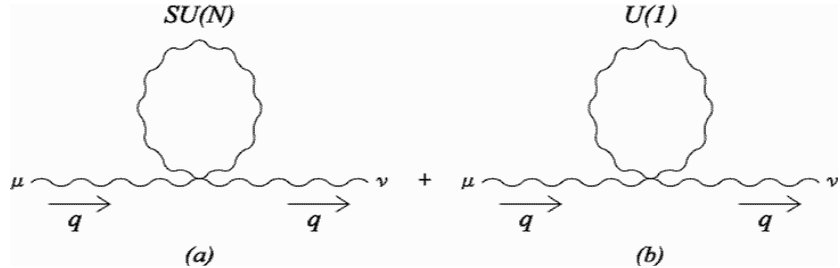}}
\vspace{1cm}
\caption{One-loop $U(1)$ gluon self-energy diagrams with $SU(N)$ (a) and
$U(1)$ (b) loops, respectively.}
\end{figure}
\beq
{\rm Fig.~8(a)}&=&
-6g^2g^{\mu\nu}
\int\!\frac{d^4l}{(2\pi)^4}
\frac 1{l^2+i\epsilon}\langle \sin^2{\frac 12(q\wedge l)}
\rangle,\nn\\[2mm]
{\rm Fig.~8(b)}&=&
-\frac 6Ng^2g^{\mu\nu}
\int\!\frac{d^4l}{(2\pi)^4}
\frac 1{l^2+i\epsilon}\langle \sin^2{\frac 12(q\wedge l)}
\rangle.
\label{eqn:4-12}
\end{eqnarray}
Both of them vanish in the new UV limit with the condition 
(\ref{eqn:2-23}).\footnote{To obtain (\ref{eqn:2-26}) form the second 
equation of (\ref{eqn:4-12}) multiply $2N$ since 
$T_0=\frac 1{\sqrt{2N}}{\boldsymbol 1}_N$.} That is, $U(1)$ decouples from 
$SU(N)$ in the one-loop approximation for the self-energy diagram.
%%%%%%%%%%%%%%%%%%%%%%%%%%%%%%%%%%%%%%%%%%%%%%%%%%%%%%%%%%
\section{Discussions} 
%%%%%%%%%%%%%%%%%%%%%%%%%%%%%%%%%%%%%%%%%%%%%%%%%%%%%%%%%%
We have presented several model calculations in the previous\cite{1)} 
and this papers that NC regularization works in non-gauge and gauge 
theories. The scenario of the method is based on the observation that, 
since UV divergence in QFT is renormalized away, the commutative limit 
of our Lorentz-invariant NCQFT must exhibit IR divergences to be 
subtracted off, if the IR limit and  the commutative limit cannot be 
distinguishable.\footnote{Long wave length `sees' the space-time in 
a coarse way, that is, in the IR limit, the space-time 
non-commutativity loses its meaning.} Indeed, one cannot discriminate
the two limits as far as one-loop self-energy diagrams
in Lorentz-invariant NCQFT are concerned. Moreover, it is important
to recognize that the two limits have invariant meanings. Lorentz 
invariance unravels the hitherto-unknown aspect of the IR/UV mixing.
\\
\ind
As remarked in I and reemphasized in \S 1 of this paper our use of 
Lorentz-invariant NCQFT as a means of the regularization in QFT is 
motivated to understand the IR/UV mixing in an invariant way. The 
elimination of the IR singularity is necessitated to make sense the 
Lorentz-invariant NCQFT quantum mechanically. There is alternative 
approach\cite{10),12)} to the Lorentz-invariant NCQED using Seiberg-Witten 
map.\cite{3)} It tries to look for small effects arising from the 
nonvanishing small value of the fundamental length $a$. In this approach 
Feynman rules in the theory are the same as those of the commutative 
fields, regarding the Lorentz-invariant NCQED as an effective field theory. 
There is no vertex factor like $V(p,q)$ as introduced in \S2.
\\
\ind
Considering this possibility we may argue that the Lorentz-invariant
NCQFT has dual roles. On one hand, it provides a kind of 
regularization by taking the new UV limit in which we let $a\to 0$.
On the other hand, we seek for new physical effects by allowing $a$
to remain finite but extremely small with only known Feynman rules 
being encountered.
\\
\ind
We have not yet checked consistency on the decoupling of $U(1)$
from $SU(N)$ since evaluation of multi-points vertices and
higher-loops are still beyond our present ability. For instance, one 
may suppose that our one-loop
calculation indicates {\it different} running of $U(1)$ and $SU(N)$ 
coupling constants, which may clash with $\ast$-gauge invariance.
On the other hand, one may also suppose that, if $SU(N)$ is not
broken as color, $U(1)$ is neither broken as $U(1)_{em}$. We shall
study these and other problems including renormalization program
in our scheme step by step.
\section*{Acknowledgements}
The author is grateful to
H. Kase and Y. Okumura for useful discussions.
\appendix
%%%%%%%%%%%%%%%%%%%%%%%%%%%%%%%%%%%%%%%%%%%%%%%%%
\section{Some mathematical formulae}
%%%%%%%%%%%%%%%%%%%%%%%%%%%%%%%%%%%%%%%%%%%%%%%%%
\indent
We collect here some mathematical formulae used in \S 2 
and \S 3 from I. For typographical reason we omit the index $E$ and work in 
Euclidean metric, $g^{\mu\nu}=-\delta^{\mu\nu}, \mu,\nu=1,2,3,4,$
with $q\cdot l=q^4l^4+{\boldsymbol q}\cdot{\boldsymbol l}$.
\\
\ind
The definition (\ref{eqn:2-16}) reads in this notation
\begin{eqnarray}
\int\!\frac{d^4l}{(2\pi)^4}
\frac{l^\mu l^\nu}{(l^2+\Delta)^2}
V^2(q,l)&=&C_1g^{\mu\nu}
+C_2q^\mu q^\nu,\\[2mm]
\int\!\frac{d^4l}{(2\pi)^4}
\frac 1{l^2+\Delta}
V^2(q,l)&=&C_3.
\label{eqn:A-2}
\end{eqnarray}
To evaluate $l$-integral it is necessary to determine the extra vertex 
factor $V(q,l)$. Since there is no guiding principle to determine the 
weight function, we employ the simplest, namely, Gaussian weight 
function:\footnote{Euclidean form of the normalization
$\int\!d^6{\btheta}w(\btheta)=(-i)^3\int\!d^6{\btheta}_Ew_E(\btheta_E)=1$ 
implies that the following $w({\bar\theta})=(-i)^3w_E(\btheta_E)$.
Our choice corresponds to positive ${\bar\alpha}=\frac 12
\btheta^{\mu\nu}\btheta_{\mu\nu}$ which is disconnected from the negative
${\bar\alpha}$.\cite{13)}}
\begin{eqnarray}
w({\bar\theta})=\frac 1{\pi^3}
e^{-b[({\bar\theta}^{41})^2+({\bar\theta}^{42})^2+({\bar\theta}^{43})^2
+({\bar\theta}^{12})^2
+{\bar\theta}^{23})^2+({\bar\theta}^{31})^2]}, b>0.
\label{eqn:A-3}
\end{eqnarray}
The extra vertex factor is then determined\cite{12)} as
\beq
V(p,l)&=&
e^{-\frac{A}2[l^2p^2-(p\cdot l)^2]},
\label{eqn:A-4}
\end{eqnarray}
where 
\beq
A=\frac {a^4}2\frac{\langle {\bar\theta}^{\,2}\rangle}{24}
\label{eqn:A-5}
\end{eqnarray}
with $\langle {\bar\theta}^{\,2}\rangle=6/b$. Since $C_{1,2}$ 
are functions of invariant $q^2$ only, we calculate the integral in 
(A$\cdot$ 1) by choosing the 4-th direction in $l$-space as pointing 
to the vector $q$ so that
\beq
q&=&(0,0,0,q),\quad l=(l^1,l^2,l^3,l^4),\nn\\[2mm]
l^4&=&l\cos{\theta_1},\quad l^3=l\sin{\theta_1}\cos{\theta_2},\nn\\[2mm]
l^2&=&l\sin{\theta_1}\sin{\theta_2}\cos{\theta_3},
\quad l^1=l\sin{\theta_1}\sin{\theta_2}\sin{\theta_3}.
\label{eqn:A-6}
\eeq
The $\mu=\nu=4$ component of (A$\cdot$ 1) is then given by
\beq
\int\!\frac{d^4l}{(2\pi)^4}
\frac{l^2\cos^2{\theta_1}}{(l^2+\Delta)^2}
e^{-Al^2q^2\sin{\theta_1}}&=&
-C_1+C_2q^2,
\label{eqn:A-7}
\eeq
while the $\mu=\nu=3$ component determines $C_1$,
\beq
\int\!\frac{d^4l}{(2\pi)^4}
\frac{l^2\sin^2{\theta_1}\cos^2{\theta_2}}{(l^2+\Delta)^2}
e^{-Aq^2l^2\sin^2{\theta_1}}&=&
-C_1.
\label{eqn:A-8}
\eeq
The $l$-integrals in  (A$\cdot$ 7) and  (A$\cdot$ 8) can easily be done 
in the spherical coordinates using Schwinger representation to yield
\beq
C_1&=&-\frac 1{32\pi^2}
\int_0^\infty\!ds 
\frac {\sqrt{s}e^{-s\Delta}}{\big(\sqrt{s+Aq^2}\big)^5},\nn\\[2mm]
C_2&=&
\frac 1{32\pi^2}A
\int_0^\infty\!ds 
\frac {e^{-s\Delta}}
{\sqrt{s}\big(\sqrt{s+Aq^2}\big)^5}.
\label{eqn:A-9}
\end{eqnarray}
On the other hand, the definition for $C_3$ leads to the result
\beq
C_3&=&
\int\!\frac{d^4l}{(2\pi)^4}
\frac 1{l^2+\Delta}
e^{-Aq^2l^2\sin^2{\theta_1}}\nn\\[2mm]
&=&\frac 1{16\pi^2}
\int_0^\infty\!ds
\frac{e^{-s\Delta}}{\sqrt{s}\big(\sqrt{s+Aq^2}\big)^3}.
\label{eqn:A-10}
\end{eqnarray}
The relation
\beq
2C_1+C_3=2C_2q^2
\label{eqn:A-11}
\eeq
follows immediately. Equations (A$\cdot$ 9), (A$\cdot$ 10) and (A$\cdot$ 11)
are reported in (\ref{eqn:2-17}) and (\ref{eqn:2-18}).
\\
\ind
The integrals considered so far are divergent as $a\to 0$ where $V(q,l)\to 1$.
This divergent behavior is transferred to IR singularity as seen from 
(A$\cdot$ 9) and (A$\cdot$ 10). It may not be uninteresting to see what 
happens for `convergent' integrals at $a=0$ ignoring Ward-Takahashi identity.
We expect that they possess no IR singularity at all and the relation 
(A$\cdot$ 11) breaks down.
\\
\ind
To be definite we consider the following `convergent' integrals
\begin{eqnarray}
\int\!\frac{d^4l}{(2\pi)^4}
\frac{l^\mu l^\nu}{(l^2+\Delta)^4}
V^2(q,l)&=&D_1g^{\mu\nu}
+D_2q^\mu q^\nu,\\[2mm]
\int\!\frac{d^4l}{(2\pi)^4}
\frac 1{(l^2+\Delta)^3}
V^2(q,l)&=&D_3.
\label{eqn:A-12}
\end{eqnarray}
We use the same function (A$\cdot$ 4) for $V(q,l)$. The result turns out to be
\beq
D_1&=&
-\frac 1{32\pi^2}\frac 1{\Gamma(4)}
\int_0^\infty\!
ds
\frac {s^{5/2}e^{-s\Delta}}
{\big(\sqrt{s+Aq^2}\big)^5},\nn\\[2mm]
D_2&=&
\frac 1{32\pi^2}\frac 1{\Gamma(4)}
A\int_0^\infty\!
ds
\frac {s^{3/2}e^{-s\Delta}}
{\big(\sqrt{s+Aq^2}\big)^5},\nn\\[2mm]
D_3&=&
\frac 1{16\pi^2}\frac 1{\Gamma(3)}
\int_0^\infty\!ds
\frac{s^{3/2}e^{-s\Delta}}
{\big(\sqrt{s+Aq^2}\big)^3}.
\label{eqn:A-13}
\end{eqnarray}
These integrals have no IR singularity because they are convergent 
at $a=0$.\footnote{The log divergence of the $s$-integral in $D_2$ 
in the commutative limit is annihilated by the factor $A$.}
Instead of the relation (A$\cdot$ 11) we find
\beq
6D_1+D_3=6D_2q^2.
\label{eqn:A-14}
\eeq
The absence of IR singularity makes the relation (A$\cdot$ 11) change into a 
different one like (A$\cdot$ 15). The transversality of the amplitudes
$\stackrel{\!\!\!(1)}{\Pi_{b,f(2)}^{\mu\nu}}(q)$
can be proven only for (\ref{eqn:2-11}) and (\ref{eqn:3-4}).
\\
\ind
There is a similar circumstance in dimensional regularization.
Although the integral
\beq
I^{\mu\nu}=
\int\!\frac{d^nl}{(2\pi)^n}
\frac{2l^\mu l^\nu-g^{\mu\nu}(l^2-\Delta)}
{(l^2-\Delta+i\epsilon)^2}
\eeq
vanishes in dimensional regularization as shown in \S 3, the integral
\beq
J^{\mu\nu}=
\int\!\frac{d^nl}{(2\pi)^n}
\frac{2l^\mu l^\nu-g^{\mu\nu}(l^2-\Delta)}
{(l^2-\Delta+i\epsilon)^4}
\eeq
does not vanish for $n\to 4$.
%%%%%%%%%%%%%%%%%%%%%%%%%%%%%%%%%%%%%%%%%%%%%%%%%
\section{Tadpole contribution to photon self-energy}
%%%%%%%%%%%%%%%%%%%%%%%%%%%%%%%%%%%%%%%%%%%%%%%%%
%
\indent
Set $\Delta V=\langle e^{-iq\wedge l}\rangle
-\langle e^{\frac i2q\wedge l}\rangle^2$.
We prove that the integral
\beq
K(q^2)=i
\int\!\frac{d^4l}
{(2\pi)^4}
\frac {\Delta V}{l^2-m^2+i\epsilon}
\label{eqn:B-1}
\eeq
vanishes in the new UV limit. Wick rotation gives
\beq
K(-q_E^2)=
\int\!\frac{d^4l_E}
{(2\pi)^4}
\frac {\Delta V_E}{l_E^2+m^2}.
\label{eqn:B-2}
\eeq
Using (A$\cdot$ 4) we have
\beq
K(-q_E^2)&=&
\int\!\frac{d^4l_E}
{(2\pi)^4}
\frac 1{l_E^2+m^2}
[e^{-2A_E(q_E^2l_E^2-(q_E\cdot l_E)^2)}
-e^{-A_E(q_E^2l_E^2-(q_E\cdot l_E)^2)}],
\label{eqn:B-3}
\eeq
which is cast into the form by (A$\cdot$ 10)
\beq
K(-q_E^2)
&=&\frac 1{16\pi^2}
\int_0^\infty\!ds\big[
\frac{e^{-sm^2}}{\sqrt{s}\big(\sqrt{s+2A_Eq_E^2}\big)^3}
-\frac{e^{-sm^2}}{\sqrt{s}\big(\sqrt{s+A_Eq_E^2}\big)^3}\big].
\label{eqn:B-4}
\end{eqnarray}
Analytic continuation back to Minkowski metric introduces the UV cutoff as
in (\ref{eqn:2-22}),
\beq
K(q^2,\Lambda^2)
&=&\frac 1{16\pi^2}
\int_0^\infty\!ds\big[
\frac{e^{-sm^2-\frac 1{s\Lambda^2}}}
{\sqrt{s}\big(\sqrt{s-2Aq^2}\big)^3}
-\frac{e^{-sm^2-\frac 1{s\Lambda^2}}}
{\sqrt{s}\big(\sqrt{s-Aq^2}\big)^3}\big].
\label{eqn:B-5}
\end{eqnarray}
We may expand the integrand with respect to $Aq^2$ and take the new 
UV limit (\ref{eqn:1-7})
to obtain
\beq
\mathop{\rm lim}_{\Lambda^2\to\infty,
a^2\to 0, \Lambda^2a^2:{\rm fixed}}\;
K(q^2,\Lambda^2)
=\mathop{\rm lim}_{\Lambda^2\to\infty,
a^2\to 0, \Lambda^2a^2:{\rm fixed}}\;
\frac 3{32\pi^2}Aq^22(m^2\Lambda^2)K_2\big(2\sqrt{m^2/\Lambda^2}\big)
\label{eqn:B-6}
\eeq
which vanishes by the condition (\ref{eqn:2-23}) since $K_2(z)\to 2/z^2$ 
as $z\to 0$. This proves that the square-bracketed term in (\ref{eqn:2-10}) 
can be neglected in the new UV limit with (\ref{eqn:2-23}).
\\
\ind
We may skip the above detailed calculation by noting that
$\Delta V$ behaves like $a^4l^2$ for small $a$.
Inserting $a^4l^2$ for $\Delta V$ in (B$\cdot$ 1) we find 
$a^4$ times a quartic divergent integral, that is, $K(q^2)$ is 
essentially given by $a^4\Lambda^4$ which vanishes by the 
condition (\ref{eqn:2-23}).\footnote{Higher terms
$a^{4n}l^{2n}$ give $a^{4n}\Lambda^{2(n+1)}$ which can be neglected
in the new UV limit.}
%%%%%%%%%%%%%%%%%%%%%%%%%%%%%%%%%%%%%%%%%%%%%%%%%
\section{Tadpole contribution to $SU(N)$ gluon self-energy}
%%%%%%%%%%%%%%%%%%%%%%%%%%%%%%%%%%%%%%%%%%%%%%%%%
Here put $\Delta'V=\langle \cos^2{\frac 12l\wedge q}\rangle
-\langle \cos{\frac 12l\wedge q}\rangle^2$.
We prove that the integral
\beq
L(q^2)=i
\int\!\frac{d^4l}
{(2\pi)^4}
\frac {\Delta' V}{l^2+i\epsilon}
\label{eqn:C-1}
\eeq
vanishes in the new UV limit. Using $\Delta'V=1-
\langle \sin^2{\frac 12l\wedge q}\rangle
-\langle \cos{\frac 12l\wedge q}\rangle^2$ and noting the fact
that (\ref{eqn:2-26}) vanishes in the new UV limit, we may replace
$\Delta'V$ in (\ref{eqn:C-1}) with $\Delta''V=1-
\langle \cos{\frac 12l\wedge q}\rangle^2$ to get
\beq
L(q^2)=i
\int\!\frac{d^4l}
{(2\pi)^4}
\frac {\Delta'' V}{l^2+i\epsilon}.
\label{eqn:C-2}
\eeq
Wick rotation and use of (\ref{eqn:A-10}) yields
\beq
L(q^2,\Lambda^2)=
\int\!\frac{d^4l_E}
{(2\pi)^4}
\frac 1{l_E^2}|_{\Lambda^2}
-\frac 1{16\pi^2}
\int_0^\infty\!ds
\frac{e^{-\frac 1{s\Lambda^2}}}
{\sqrt{s}\big(\sqrt{s-Aq^2}\big)^3},
\label{eqn:C-3}
\eeq
where we regulated the integrals.
This goes like $a^4\Lambda^4$ in the new UV limit and can be neglected
if we impose the condition (\ref{eqn:2-23}). This allows us to evaluate
the tadpole diagram for $S(N)$ gluon self-energy by replacing the
extra vertex factor $\langle \cos^2{\frac 12(l\wedge q)}\rangle$
with $\langle \cos{\frac 12(l\wedge q)}\rangle^2$ as done in \S 4.
Similar proof goes through for all other tadpole diagrams.

\end{document}